*Title:* **Recent Developments of the Cascade-Exciton Model of Nuclear Reactions**

*Author(s):* Stepan G. MASHNIK and Arnold J. SIERK

# Recent Developments of the Cascade-Exciton Model of Nuclear Reactions

Stepan G. MASHNIK* and Arnold J. SIERK

*T-16, Theoretical Division, Los Alamos National Laboratory, Los Alamos, NM 87545, USA*

Recent developments of the Cascade-Exciton Model (CEM) of nuclear reactions are described. The improved cascade-exciton model as implemented in the code CEM97 differs from the CEM95 version by incorporating new approximations for the elementary cross sections used in the cascade, using more precise values for nuclear masses and pairing energies, using corrected systematics for the level-density parameters, and several other refinements. We have improved algorithms used in many subroutines, decreasing the computing time by up to a factor of 6 for heavy targets. We describe a number of further improvements and changes to CEM97, motivated by new data on isotope production measured at GSI. This leads us to CEM2k, a new version of the CEM code. CEM2k has a longer cascade stage, less preequilibrium emission, and evaporation from more highly excited compound nuclei compared to earlier versions. CEM2k also has other improvements and allows us to better model neutron, radionuclide, and gas production in ATW spallation targets. The increased accuracy and predictive power of the code CEM2k are shown by several examples. Further necessary work is outlined.

*KEYWORDS: Intranuclear cascade, preequilibrium, evaporation, and fission reactions, Monte Carlo simulations, cascade-exciton model, particle spectra, spallation and fission cross sections, GSI data*

## I. Introduction

The design of a hybrid reactor system driven with a high current proton accelerator requires information about residual nuclides that are produced by high energy protons and secondary neutrons interacting in the target and in structural materials. It is both physically and economically impossible to measure all necessary data, which is why reliable models and codes are needed. A model with a good predictive power both for the spectra of emitted particles and residual nuclide yields is the Cascade-Exciton Model (CEM) of nuclear reactions.[1] A more modern version of the CEM is implemented in the code CEM95,[2] which is available from the NEA/OECD, Paris. Following an increased interest in intermediate-energy nuclear data inspired by such projects as Accelerator Transmutation of Wastes (ATW), Accelerator Production of Tritium (APT), and others, we developed a newer version of the cascade-exciton model, CEM97.[3,4] CEM97 has a number of improved physics features and due to significant algorithmic improvements is several times faster than CEM95. It has been incorporated into the transport code system MCNPX.[5]

Recent GSI measurements of interactions of $^{208}$Pb[6,7] and $^{238}$U[8] at 1 GeV/nucleon and $^{197}$Au at 800 MeV/nucleon[9] with liquid $^1$H provide a very rich set of cross sections for production of practically all possible isotopes from these reactions in a "pure" form, *i.e.*, individual cross sections from a specific given bombarding isotope (or target isotope, when considering reactions in the usual kinematics, p + A). Such cross sections are much easier to compare to models than the "camouflaged" data from $\gamma$-spectrometry measurements. These are often obtained only for a natural composition of isotopes in a target and are mainly for cumulative production, where measured cross sections contain contributions not only from direct production of a given isotope, but also from all its decay-chain precursors. Analysis of these new data has motivated us to further improve the CEM and to develop a version of a new code, CEM2k,[10] still under development.

## II. Results and Discussion

First, we review the basis of the CEM and the main differences between the improved cascade-exciton model code CEM97[3,4] and its precursor, CEM95.[2] The CEM assumes that reactions occur in three stages. The first stage is the IntraNuclear Cascade (INC) in which primary particles can be re-scattered and produce secondary particles several times prior to absorption by or escape from the nucleus. The excited residual nucleus remaining after the cascade determines the particle-hole configuration that is the starting point for the preequilibrium stage of the reaction. The subsequent relaxation of the nuclear excitation is treated in terms of an improved Modified Exciton Model (MEM) of preequilibrium decay followed by the equilibrium evaporative final stage of the reaction. Generally, all three stages contribute to experimentally measured outcomes.

An important ingredient of the CEM is the criterion for transition from the intranuclear cascade to the preequilibrium model. In conventional cascade-evaporation models (like ISABEL and Bertini's INC used in LAHET[11]) fast particles are tracked down to some minimal energy, the cutoff energy $T_{cut}$ (or to a cutoff time, in the "time-like" INC models, like the Liege INC[12]) which is usually about 7–10 MeV above the interior nuclear potential, below which particles are considered to be absorbed.

The CEM uses a different criterion to decide when a particle is considered to have left the cascade. An effective local optical absorptive potential $W_{opt.\ mod.}(r)$ is defined from the local interaction cross section of the particle, including Pauli-blocking effects. This imaginary potential is compared to one defined by a phenomenological global optical model $W_{opt.\ exp.}(r)$. We characterize the degree of similarity or dif-

* Corresponding author, Tel. +1-505-667-9946, Fax. +1-505-667-1931, E-mail: mashnik@t2y.lanl.gov

ference of these imaginary potentials by the parameter

$$\mathcal{P} = |(W_{opt.\ mod.} - W_{opt.\ exp.})/W_{opt.\ exp.}|.$$

When $\mathcal{P}$ increases above an empirically chosen value, the particle leaves the cascade, and is then considered to be an exciton. Both CEM95 and CEM97 use the fixed value $\mathcal{P} = 0.3$. With this value, we find the cascade stage of the CEM is generally shorter than that in other cascade models.

The transition from the preequilibrium stage to the evaporation stage occurs when the probability of nuclear transitions changing the number of excitons $n$ with $\Delta n = +2$ becomes equal to the probability of transitions in the opposite direction, with $\Delta n = -2$, i.e., when the exciton model predicts an equilibration has been established in the nucleus.

The improved cascade-exciton model in the code CEM97 differs from the CEM95 version by incorporating new approximations for the elementary cross sections used in the cascade, using more precise values for nuclear masses and pairing energies, using corrected systematics for the level-density parameters, improving the approximation for the pion "binding energy", $V_\pi$, adjusting the cross sections for pion absorption on quasi-deuteron pairs inside a nucleus, considering the effects on cascade particles of refractions and reflections from the nuclear potential, allowing for nuclear transparency of pions, including the Pauli principle in the preequilibrium calculation, and implementing significant refinements and improvements in the algorithms of many subroutines, decreasing the computing time by up to a factor of 6 for heavy nuclei, which is very important when performing practical simulations with transport codes like MCNPX. On the whole, this set of improvements leads to a better description of particle spectra and yields of residual nuclei and a better agreement with available data for a variety of reactions. Details and examples with some results may be found in.[3,13]

We also made a number of refinements in the calculation of the fission channel, as described in.[4,14,15] The original version of CEM95[2] incorporates a user-selected level-density-parameter formula that is applied to all decay channels of an excited nucleus except for the fission channel. For fission, the level-density parameter at the saddle point $a_f$ is calculated using an analogous parameter for the neutron emission channel, $a_n$ and a constant ratio $a_f/a_n$ which serves as a fitting parameter of the model. Thus the shell-effect influence on the level density in the neutron emission channel is automatically conveyed to the level density at the saddle point. But we expect that shell corrections at the saddle point should bear no relation to those at the ground state, due to the large saddle-point deformation, and the consequent different microscopic level structure near the Fermi surface.

We have performed calculations for neutron- and proton-induced reactions on several targets, focusing on $^{208}$Pb and $^{209}$Bi, with the parameter $a_f$ being *energy-independent*, which is the same as ignoring the ground-state shell effect on the level density at the saddle point (see[14]). An example of these calculations with the original and an improved version of CEM as described in[14] is shown in **Fig. 1**. There is a better description of the experimental fission cross sections in comparison with the original version. Unfortunately, there remains one empirical parameter, $B_s$ (the normalized surface area), but the energy dependence is much better reproduced. Similar improvements were obtained for fission cross sections induced by intermediate energy $\gamma$ and $\pi^-$ on Sn, Au, Bi, and $^{238}$U.[4,15]

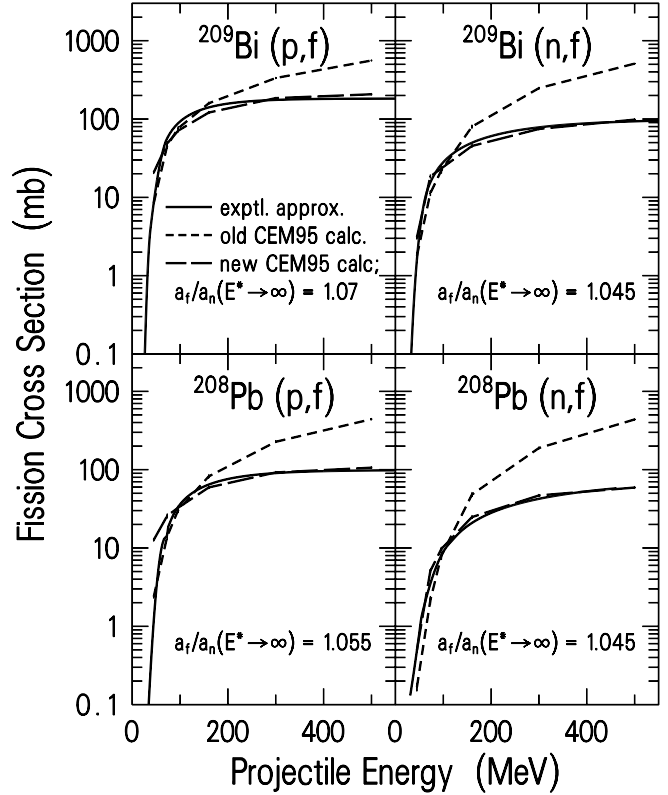

**Fig. 1** Comparison between the experimental data and calculations of the cross sections for the reactions $^{209}$Bi(p,f), $^{209}$Bi(n,f), $^{208}$Pb(p,f), and $^{208}$Pb(n,f) using the original and modified versions of CEM95. The solid lines represent the parametrization by Prokofiev[14,16] of the experimental data. The short-dashed lines show the original CEM95 results and the long-dashed lines, the results with a fitted energy-independent value for $a_f$.

Besides this modification of the CEM95 code introduced especially for a better description of fission cross sections, we have been further improving the CEM,[3,4] striving for a model capable of predicting different characteristics of nuclear reactions for arbitrary targets with a wide range of incident energies. Modifications made for a better description of the preequilibrium, evaporative, and cascade stages will also affect the fission channel. We have incorporated into the CEM the updated experimental atomic mass table by Audi and Wapstra,[17] the nuclear ground-state masses (where data does not exist), deformations, and shell corrections by Möller et al.,[18] and the pairing energy shifts from Möller, Nix, and Kratz[19] into the level-density formula. In addition, we have derived a corrected systematics for the level-density parameters using the Ignatyuk expression,[20] with coefficients fitted to the data analyzed by Iljinov et al.[21] (we discovered that Iljinov et

*al.* used $11/\sqrt{A}$ for the pairing energies $\Delta$ (see Eq. (3) in[4]) in deriving their level-density systematics instead of the value of $12/\sqrt{A}$ stated in[21] ). We also derived additional semiempirical level-density-parameter systematics using the Möller *et al.*[18] ground-state microscopic corrections, both with and without the Möller, Nix, and Kratz[19] pairing gaps, and introduced into the CEM a new empirical relation to take into account the excitation-energy dependence of the ground-state shell correction in the calculation of fission barriers (see[4]).

As mentioned in the Introduction, analysis of the recent GSI measurements[6–9] has motivated us to further improve the CEM. The authors of the GSI measurements performed a comparison of their data to several codes, including LAHET,[11] YIELDX,[23] ISABLA (ISABEL INC code from LAHET followed by ABLA[24] evaporation code), CASCADO,[25] and the Liege INC by Cugnon,[12] and encountered serious problems; none of these codes were able to accurately describe all their measurements. Most of the calculated distributions of isotopes produced as a function of neutron number were shifted toward larger masses as compared to the data. While in some disagreement with the measurements, the Liege INC and the CASCADO codes provide a better agreement with the data than LAHET, ISABLA, and YIELDX do. Being aware of this situation with the GSI data, we decided to consider them ourselves, leading to the development of CEM2k.

First, we calculated the $^{208}$Pb GSI reaction[6] with the standard versions of CEM95 and CEM97 and determined[10] that though CEM95 describes quite well production of several heavy isotopes near the target (we calculate p + $^{208}$Pb; therefore $^{208}$Pb is a target, not a projectile as in the GSI measurements), it does not reproduce correctly the cross sections for lighter isotopes in the deep spallation region. The disagreement increases with increasing distance from the target, and all calculated curves are shifted to the heavy-mass direction, just as was obtained by the authors of the GSI measurements. The results of the CEM97 code are very similar to those of CEM95 (see a figure with CEM97 results in[26]). Later on, we performed an extensive set of calculations of the same data with several more codes (HETC,[27] LAHET[11] with both ISABEL and Bertini options, CASCADE,[28] CASCADE/INPE,[29] INUCL,[30] and YIELDX[23]) and got very similar results:[26] all codes disagree with the data in the deep spallation region, the disagreement increases as one moves away from the target, and all calculated curves are shifted in the heavy-mass direction.

This means that for a given final element (Z), all models predict emission of too few neutrons. Most of the neutrons are emitted at the final (evaporation) stage of a reaction. One way to increase the number of emitted neutrons would be to increase the evaporative part of a reaction. In the CEM, this might be done in two different ways: the first would be to have a shorter preequilibrium stage, so that more excitation energy remains available for the following evaporation; the second would be to have a longer cascade stage, so that after the cascade, less exciton energy is available for the preequilibrium stage, so fewer energetic preequilibrium particles are emitted,

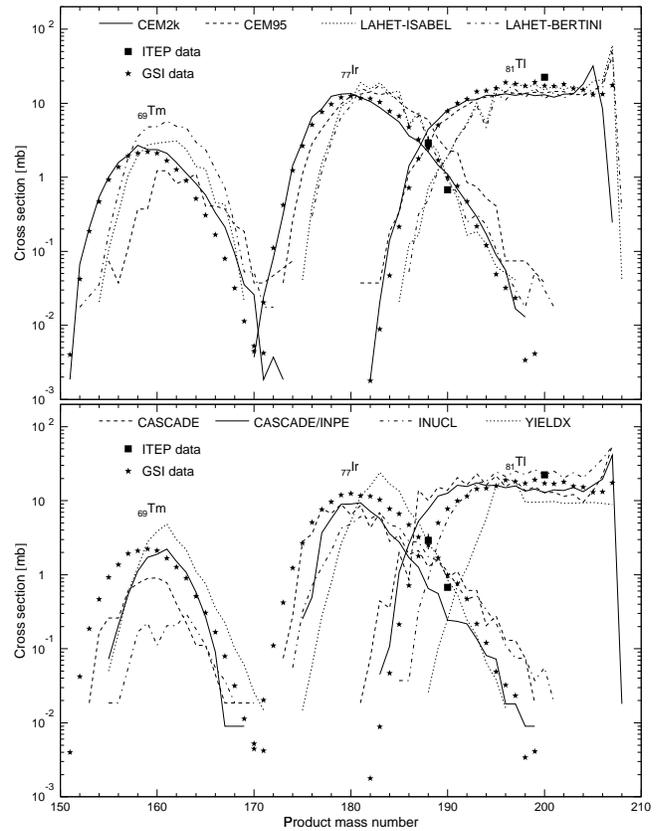

**Fig. 2** Mass distributions for independent production of Tm, Ir, and Tl isotopes from 1 GeV protons colliding with $^{208}$Pb. Stars are GSI measurements,[6] while squares show ITEP data.[31] The calculations are identified as: CEM2k—our results, CEM95—Ref.,[2] LAHET-ISABEL—Ref.,[11] LAHET-Bertini—Ref.,[11] CASCADE/INPE—Ref.,[29] CASCADE—Ref.,[28] INUCL—Ref.,[30] and YIELDX—Ref.[23]

leaving more excitation energy for the evaporative stage.

One easy way to shorten the preequilibrium stage of a reaction in CEM is to arbitrarily allow only transitions that increase the number of excitons, $\Delta n = +2$, *i.e.*, only allow the evolution of a nucleus toward the compound nucleus. In this case, the time of the equilibration will be shorter and fewer preequilibrium particles will be emitted, leaving more excitation energy for the evaporation. Such an approach is used by some other exciton models, for instance, by the Multistage Preequilibrium Model (MPM) used in LAHET.[11] Calculations using this modification to CEM97 (see Fig. 2d in[10]) provide a shift of the calculated curves in the right direction, but only slightly improve agreement with the GSI data.

A second method of increasing evaporation is to enlarge the cascade part of a reaction; we may either enlarge the parameter $\mathcal{P}$ or remove it completely and resort to a cutoff energy $T_{cut}$, as is done in other INC models. Calculations have shown that a reasonable increase of $\mathcal{P}$ doesn't solve the problem. However, if we do not use $\mathcal{P}$, but instead use a cutoff energy of $T_{cut} = 1$ MeV for incident energies above the pion production threshold, the code agrees with the GSI data significantly better (see Fig. 2e in[10]). Using both these conditions leads to results that describe the p + $^{208}$Pb GSI data very well. We call this approach CEM2k. An example of CEM2k results for the yield of Tm, Ir, and Tl isotopes from p + $^{208}$Pb interactions compared to the GSI[6] and ITEP[31] data and with predictions by CEM95, LAHET-ISABEL, LAHET-Bertini, CASCADE, CASCADE/INPE, INUCL, and YIELDX is shown in **Fig. 2**. Similar comparisons for more isotopes may be found in.[10,26,31] We find that CEM2k agrees best with these GSI (and ITEP) data of the codes tested here and in.[10,26,31]

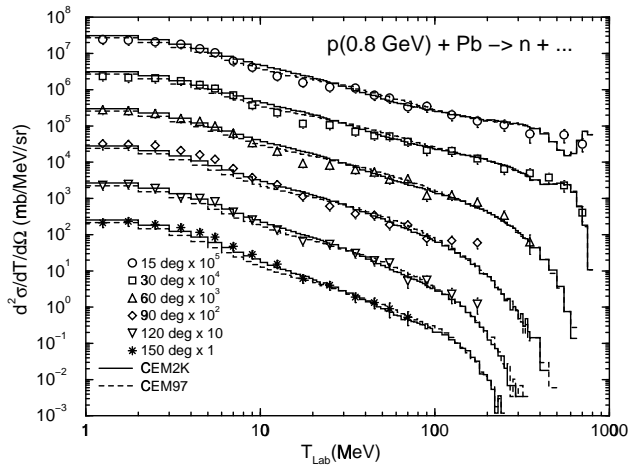

Fig. 3 Comparison of measured[32] double differential cross sections of neutrons from 0.8 GeV protons on Pb with CEM2k and CEM97 calculations.

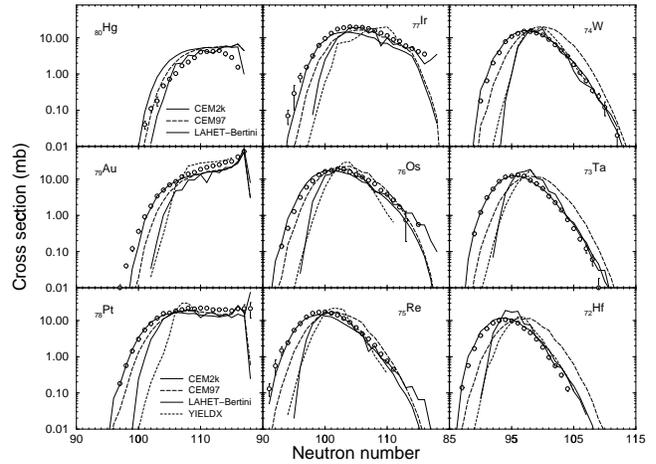

Fig. 4 Isotopic distribution of spallation products from the reaction $^{197}$Au + p at 800 $A$ MeV from mercury to hafnium. Open circles are the GSI data,[9] CEM2k (thick solid curves) and CEM97 (thick dashed curves) are our present calculations, LAHET-Bertini (thin solid curves) and YIELDX (thin dashed curves) are calculations from.[9]

Finding a good agreement of CEM2k with the isotope production, we wish to see how well it describes spectra of secondary particles in comparison with CEM97. **Figure 3** shows examples of neutron spectra from interactions of protons with the same target, $^{208}$Pb at 0.8 GeV (we do not know of measurements of spectra at 1 GeV, the energy of the isotope-production data). We see that CEM2k describes spectra of secondary neutrons comparably to CEM97, even possibly a little better at larger angles. Similar results[10] hold for 1.5 GeV protons on $^{208}$Pb.[32] So this preliminary version of CEM2k, describes both the GSI data from $^{208}$Pb interactions with p at 1 GeV/nucleon and the spectra of emitted neutrons from p+$^{208}$Pb at 0.8 and 1.5 GeV better than its precursor CEM97.

We use CEM2k as fixed from our analysis of the $^{208}$Pb data[6,7] without further modifications to calculate the $^{197}$Au[9] and $^{238}$U[8] GSI measurements. An example of the yield of several isotopes from $^{197}$Au calculated by CEM2k is shown in **Fig. 4** together with standard CEM97 predictions and calculations by LAHET-Bertini and YIELDX codes from.[9] We see that just as in the case of the $^{208}$Pb data, CEM2k agrees best with the $^{197}$Au data in the spallation region compared to the other codes tested here. Several more results for $^{197}$Au and $^{238}$U and their detailed discussion may be found in.[10]

Besides the changes to CEM97 mentioned above, we also made a number of other improvements and refinements, such as imposing momentum-energy conservation for each simulated event (the Monte Carlo algorithm previously used in CEM provides momentum-energy conservation only statistically, on the average, but not exactly for the cascade stage of each event); using real binding energies for nucleons at the cascade stage instead of the approximation of a constant separation energy of 7 MeV used in previous versions of the CEM; using reduced masses of particles in the calculation of their emission widths instead of using the approximation of no recoil used previously. In **Figs. 5 and 6** we show that these refinements (which involve no further parameter fitting), while improve slightly the agreement with the GSI measurements (and other data on medium and heavy targets), are crucial for light targets, especially when calculating $^4$He and other fragment emission from light nuclei. This is especially important for applications were gas production is calculated.

Another improvement important for applications is better representing the total reaction cross section. Previous versions of CEM calculate the total reaction cross section, $\sigma_{in}$ (just like many other INC-type models) using the geometrical cross section, $\sigma_{geom}$, and the number of inelastic, $N_{in}$, and elastic, $N_{el}$, simulated events, namely: $\sigma_{in} = \sigma_{geom} N_{in}/(N_{in} + N_{el})$. This approach provides a good agreement with available data at incident energies above about 100 MeV, but is not reliable at lower bombarding energies. To address this problem, we have incorporated into CEM2k the NASA systematics by Tripathi *et al.*[34] for all incident protons and neutrons with energies above the maximum in the NASA reaction cross sections, and the Kalbach systematics[35] for neutrons of lower energy. As shown in **Fig. 7**, we can describe much better with CEM2k the total reaction cross sections (and correspondingly any other partial cross sections) for n- and p-induced reactions, especially at energies below about 100 MeV.

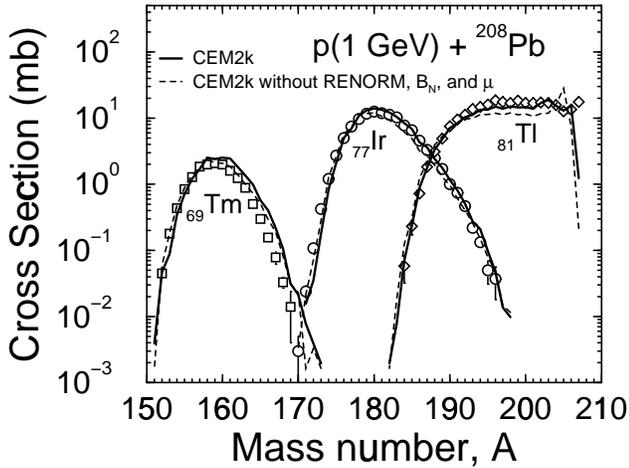

**Fig. 5** An example of how taking into account the real binding energy of the cascade nucleons, the reduced masses in calculating the widths of preequilibrium and evaporative particles, and imposing momentum-energy conservation for each cascade event in CEM2k affect the yield of Tm, Ir, and Tl nuclides from p(1 GeV) + Pb reactions; experimental data (symbols) are from GSI measurements.[6]

## III. Further Work

CEM2k is still under development. For instance, we hope to better describe complex-particle and light-fragment emission, a problem related to the poor approximations for inverse cross sections we still use in the CEM. We have compiled all experimental data and phenomenological systematics on inverse cross sections we could find and have developed a universal routine, HYBRID, which gives inverse cross sections of nucleons, complex particles up to $^4$He, as well as of heavier fragments, based on.[34, 35] We also developed algorithms allowing us to use arbitrary approximations for inverse cross sections when calculating the widths and kinetic energies of emitted particles and fragments, and developed as well a routine to calculate coalescence of complex particles from emitted fast cascade nucleons. We need to work more on this to decide what should be the heaviest fragment to be considered as produced via evaporation from a compound nucleus formed in a reaction and to fix all parameters. For this, we need to analyze a lot of more available data with CEM2k, especially at lower incident energies.

Our work on fission in CEM is also not finished. For instance, we are not satisfied with the situation that in the improved versions of the CEM we still have an additional input parameter to describe fission cross sections: either $B_s$, in the approach[14] illustrated in Fig. 1, or $a_f/a_n$, in our later version.[4] In addition, currently, CEM has no model of fission-fragment formation. Therefore, results on nuclide yields from CEM95, CEM97, or CEM2k shown here reflect only the contribution to the total yields of the nuclides from deep spallation processes of successive emission of particles, but do not contain fission products or the contributions to spectra from evaporation from them. To be able to describe nuclide pro-duction in the fission region, currently the CEM has to be supplemented by a model of fission (*e.g.*, in the transport code MCNPX,[5] where CEM97 and CEM2k are used, they are supplemented by the RAL fission model[38]).

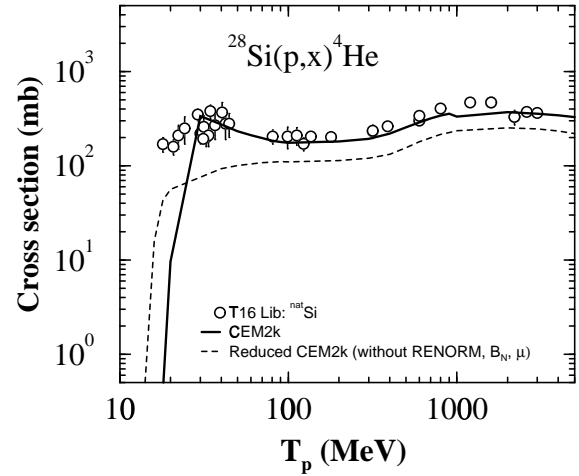

**Fig. 6** An example of how taking into account the effects discussed in previous figure affect the $^{28}$Si(p,x)$^4$He excitation function (gas production). Data (circles) are from an extended version of our compilation of experimental cross sections.[33]

When we complete CEM2k to a reasonable level, we plan to incorporate it into LAHET[11] and to replace the present version in MCNPX[5] and replace CEM95 in the MARS code system.[39]

## Acknowledgment

We thank Dr. B. Mustapha for sending us the GSI data and Dr. S. Chiba and Prof. K. Ishibashi for suppling us with their neutron spectra. We are grateful to Drs. M. B. Chadwick, N. V. Mokhov, R. E. Prael, D. D. Strottman, and L. S. Waters, for useful discussions, interest, and support. This study was supported by the U. S. Department of Energy.

## References

1) K. K. Gudima, S. G. Mashnik, and V. D. Toneev, *Nucl. Phys.* **A401**, 329 (1983).
2) S. G. Mashnik, *User Manual for the Code CEM95*, JINR, Dubna (1995); OECD NEA Data Bank, Paris, France (1995); http://www.nea.fr/abs/html/iaea1247.html.
3) S. G. Mashnik and A. J. Sierk, "Improved cascade-exciton model of nuclear reactions," *Proc. SARE4, September 14–16, 1998, Knoxville, TN, USA* (ORNL, USA, 1999) pp. 29–51; E-print *nucl-th/9812069*.
4) A. J. Sierk and S. G. Mashnik, "Modeling Fission in the Cascade-Exciton Model," *Proc. SARE4, September 14–16, 1998, Knoxville, TN, USA* (ORNL, USA, 1999) pp. 53–67; E-print *nucl-th/9812070*.
5) *MCNPX$^{TM}$ User's Manual, Version 2.1.5*, L. S. Waters, Ed., Los Alamos National Laboratory Report TPO-E83-G-UG-X-00001, Revision 0 (November 14, 1999).
6) W. Wlazlo *et al.,Phys. Rev. Lett.* **84**, 5736 (2000).
7) T. Enqvist *et al., Nucl. Phys.* **A686**, 481 (2001).
8) J. Taieb *et al.,*"Measurement of $^{238}$U Spallation Product

Cross Sections at 1 GeV per Nucleon," *Proc. RNB-5, Divanne les Bains, France, Oct. 3–8, 2000)*; http://www-wnt.gsi.de/kschmidt/publica.htm.
9) F. Rejmund *et al.*, *Nucl. Phys.* **A683**, 540 (2001); J. Benlliure *et al.*, *Nucl. Phys.* **A683**, 513 (2001).
10) S. G. Mashnik and A. J. Sierk, "CEM2k—recent developments in CEM," *Proc. AccApp00, November 13–15, 2000, Washington, DC, USA*, (American Nuclear Society, La Grange Park, IL, USA, 2001) pp. 328–341; http://xxx.lanl.gov/ps/nucl-th/0011064.
11) R. E. Prael and H. Lichtenstein, *User guide to LCS: The LAHET Code System*, LANL Report No. LA-UR-89-3014, Los Alamos (1989).
12) J. Cugnon, C. Volant, and S. Vuillier, *Nucl. Phys.* **A620**, 475 (1997).
13) S. G. Mashnik, R. J. Peterson, A. J. Sierk, and M. R. Braunstein, *Phys. Rev. C* **61**, 034601 (2000).
14) A. V. Prokofiev, S. G. Mashnik, and A. J. Sierk, *Nucl. Sci. Eng.* **131**, 78 (1999); E-print *nucl-th/9802027*.
15) R. J. Peterson *et al.*, *Eur. Phys. J.* **A10**, 69 (2001).
16) A. V. Prokofiev, *Nucl. Instr. Meth. Phys. Res.* **A463**, 557 (2001).
17) G. Audi and A. H. Wapstra, *Nucl. Phys.*, **A565**, 1 (1993).
18) P. Möller, J. R. Nix, W. D. Myers, and W. J. Swiatecki, *Atomic Data and Nuclear Data Tables*, **59**, 185 (1995).
19) P. Möller, J. R. Nix, and K.-L. Kratz, *Atomic Data and Nuclear Data Tables*, **66**, 131 (1997).
20) A. V. Ignatyuk *et al.*, *Yad. Fiz.*, **21**, 1185 (1975).
21) A. S. Iljinov *et al.*, *Nucl. Phys.*, **A543**, 517 (1992).
22) A. J. Sierk, *Phys. Rev. C*, **33**, 2039 (1986).
23) R. Silberberg, C. H. Tsao, and A. F. Barghouty, *Astrophys. J.*, **501**, 911 (1998).
24) J.-J. Gaimard and K.-H. Schmidt, *Nucl. Phys.*, **A531**, 709 (1991).
25) A. V. Ignatyuk, N. T. Kulagin, V. P. Lunev, and K.-H. Schmidt, "Analysis of Spallation Residues within the Intranuclear Cascade Model," *Proc. XV Workshop on Phys. of Nucl. Fission, Oct. 3–6, 2000, Obninsk, Russia*; http://www-wnt.gsi.de/kschmidt/publica.htm.
26) S. G. Mashnik *et al.*, "Benchmarking ten codes against the recent GSI measurements of the nuclide yields from $^{208}$Pb, $^{197}$Au, and $^{238}$U + p reactions at 1 GeV/nucleon," paper #90 in these Proceedings.
27) T. W. Armstrong and K. C. Chandler, *Nucl. Sci. Eng.* **49**, 110 (1972) and references therein.
28) V. S. Barashenkov *et al.*, JINR Report R2-85-173, Dubna, 1985; V. S. Barashenkov *et al.*, *Yad. Fiz.* **39**, 1133 (1984) [*Sov. J. Nucl. Phys.* **39**, 715 (1984)]; V. S. Barashenkov *et al.*, *Nucl. Phys.* **A338**, 413 (1980).
29) V. S. Barashenkov *et al.*, *Atomnaya Energiya* **87**, 283 (1999) [*Atomic Energy* **87**, 742 (1999)].
30) A. A. Sibirtsev *et al.*, ITEP Preprint No. ITEP-129, Moscow, 1985; N. V. Stepanov, ITEP Preprint No. ITEP-81, Moscow, 1987; N. V. Stepanov, ITEP Preprint No. ITEP-55-88, Moscow, 1988 (in Russian).
31) Yu. E. Titarenko *et al.*, LANL Report LA-UR-00-4779, Los Alamos (2000), E-print *nucl-th/0011083*, submitted to *Phys. Rev. C*.
32) K. Ishibashi *et al.*, *J. Nucl. Sci. Techn.*, **34**, 529 (1997).
33) S. G. Mashnik, A. J. Sierk, K. A. Van Riper, and W. B. Wilson, "Production and Validation of Isotope Production Cross Section Libraries for Neutrons and Protons to 1.7 GeV," *Proc. SARE4, Knoxville, TN, USA, September 14–16, 1998*, ORNL (1999)

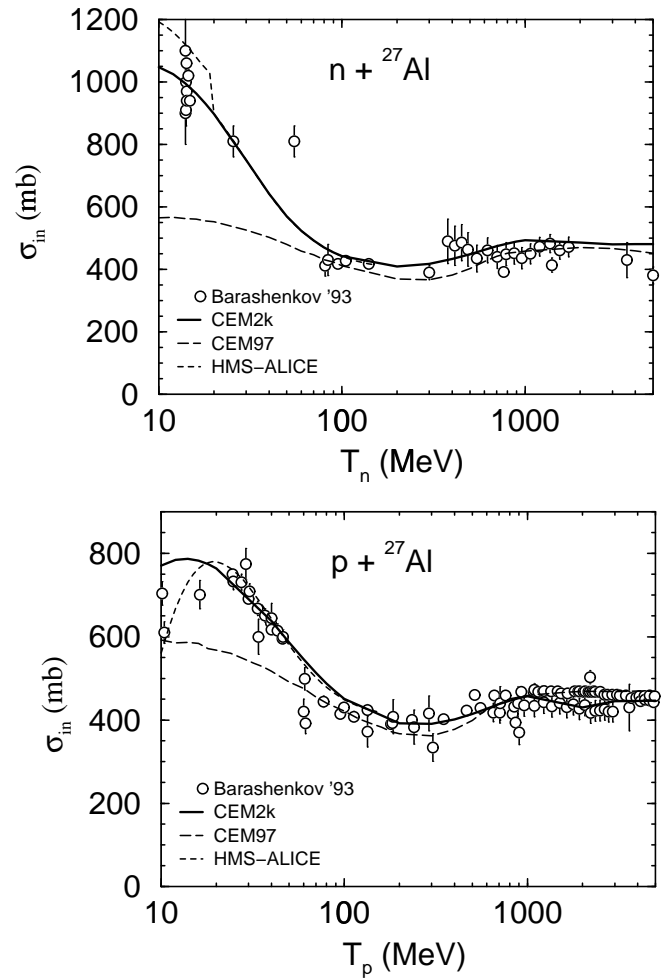

**Fig. 7** Total reaction cross sections for n- and p-induced reactions on Al calculated by CEM2k and CEM97 compared with experimental data compiled by Barashenkov[36] and calculations from the HMS-ALICE code.[37]

pp. 151–1627; E-print *nucl-th/9812071*.
34) R. K. Tripathi, F. A. Cucinotta, and J. W. Wilson, *Nucl. Instr. Meth. Phys. Res.*, **B117**, 347 (1996).
35) C. Kalbach, *J. Phys. G: Nucl. Part. Phys.*, **24**, 847 (1998).
36) V. S. Barashenkov, *Cross Sections of Interactions of Particles and Nuclei with Nuclei*, JINR, Dubna, Russia (1993); tabulated data available from NEA Data Bank at http://www.nea.fr/html/dbdata/bara.html.
37) M. Blann, *Phys. Rev. C*, **54**, 1341 (1996); M. Blann and M. B. Chadwick, *Phys. Rev. C*, **57**, 233 (1998).
38) F. Atchison, "Spallation and fission in heavy metal nuclei under medium energy proton bombardment," Jul-Conf-34, Kernforschungsanlage Julich GmbH (January 1980).
39) N. V. Mokhov, S. I. Striganov, A. Van Ginneken, S. G. Mashnik, A. J. Sierk, and J. Ranft, "MARS Code Developments," *Proc. SARE4, Knoxville, TN, USA, September 14–16, 1998*, ORNL (1999) pp. 87–997; E-print *nucl-th/9812038*.